\documentclass[runningheads]{svmult}

\usepackage{makeidx}   
\usepackage{graphicx}  
\usepackage{subeqnar}  
\usepackage{multicol}  
\usepackage{physprbb}  
\makeindex             

\begin{document}
\title*{The Starburst-AGN Connection}
\toctitle{The Starburst-AGN Connection}
%
%
\titlerunning{Starburst-AGN Connection}
%
\author{Sylvain Veilleux}
\authorrunning{Sylvain Veilleux}
%
%
\institute{Department of Astronomy, University of Maryland, College Park, MD 20742 USA}

\maketitle              

\begin{abstract}
The issue of a starburst-AGN connection in local and distant galaxies
is relevant for understanding galaxy formation and evolution, the star
formation and metal enrichment history of the universe, the origin of
the extragalactic background at low and high energies, and the origin
of nuclear activity in galaxies.  Here I review some of the
observational evidence recently brought forward in favor of a
connection between the starburst and AGN phenomena. I conclude by
raising a number of questions concerning the exact nature of this
connection.
\end{abstract}

\section{Introduction}

Since the focus of this conference is the starburst phenomenon, many
of us would rather ``sweep AGN activity under the rug'' and only
consider star formation. This would be a mistake!  There is growing
evidence that intense star formation and nuclear activity often come
hand in hand. The apparent correlation between the mass of dormant
black holes at the centers of nearby galaxies and the mass of their
spheroids (e.g., Kormendy \& Richstone 1995; Faber et al. 1997;
Magorrian et al. 1998; Gebhardt et al. 2000) suggests a direct link
between the formation of spheroids and the growth of central black
holes. Since a starburst is a natural consequence of the dissipative
gaseous processes associated with spheroid formation (e.g., Barnes \&
Hernquist 1991; Kormendy \& Sanders 1992; Mihos \& Hernquist 1994), a
starburst-AGN connection dating back to the early universe is
implied by these results.  Closer to us, the presence of
circumnuclear starbursts in an increasing number of local AGNs
(discussed in more detail in \S 2) also suggests a connection between
the starburst and AGN phenomena.

This possible starburst-AGN connection has direct bearings on our
understanding of the early universe. A large contribution from
unsuspected (hidden) AGNs would complicate the deduction of the star
formation history of the universe from galaxy luminosity functions
(e.g., Blain et al. 1999).  This in turn may change our current views
on the history of metal enrichment and the importance of feedback
processes in the early universe (e.g., Franx et al. 1997; Pettini et
al. 2000; Ferrara \& Tolstoy 2000).  Similarly, a correction needs to
be applied to account for obscured AGNs as contributors to the
far-infrared extragalactic background.  Recent X-ray observations with
{\em Chandra} have added substantially to the debate.  The discovery
that some fraction of the X-ray background appears to be produced by a
population of heavily obscured AGNs (e.g., Mushotzky et al. 2000;
Barger et al. 2000), objects which have been largely missed in optical
surveys due to extremely heavy obscuration, has clearly further
increased the importance of studies of the starburst-AGN connection in
distant infrared-selected galaxies.

The rest of this paper is organized as follows. In \S 2, I discuss
recent ($<$ 5 years approx.) results which appear to favor a starburst-AGN
connection. This discussion is not meant to be a exhaustive (or even
impartial) review of the recent literature on this subject; it is only
meant to illustrate some of the best cases where a starburst-AGN
connection indeed appears to exist. In the last section (\S 3), I
raise several questions regarding the nature of this starburst-AGN
connection.

\section{Evidence for a Starburst-AGN Connection}

Since the triggering mechanism for AGN activity probably depends on
the luminosity of the AGN, I make a distinction in the following
discussion between the nearby, low-luminosity Seyferts and
Fanaroff-Riley type I (FR I) radio galaxies and the more distant and
powerful quasars, Fanaroff-Riley type II (FR II) radio galaxies, and
ultraluminous infrared galaxies (ULIRGs; log~[L$_{\rm ir}$/L$_\odot$]
$\geq$ 12 by definition).

\subsection{Low-Luminosity Regime: Seyferts, FR I radio galaxies}

The fueling of AGNs requires mass accretion rates dM/dt $\approx$ 1.7
(0.1/$\epsilon$)(L/10$^{46}$ ergs s$^{-1}$) M$_\odot$ yr$^{-1}$, where
$\epsilon$ is the mass-to-energy conversion efficiency. A modest
accretion rate of order $\sim$ 0.01 M$_\odot$ yr$^{-1}$ is therefore
sufficient to power a Seyfert galaxy like NGC~1068. Only a small
fraction of the total gas content of a typical host galaxy is
therefore necessary for the fueling of these low-luminosity AGNs.  A
broad range of mechanisms including intrinsic processes (e.g., stellar
winds and collisions; dynamical friction of giant molecular clouds
againts stars; nuclear bars or spirals produced by gravitational
instabilities in the disk) and external processes (e.g., ``minor''
galaxy interactions or mergers) may be at work in these objects.
While a detailed discussion of these processes is beyond the scope of
the present paper (see Combes 2000 for a recent review of the
subject), suffice it to say that there is little or no observational
evidence for Seyfert nuclei to occur preferentially in barred systems
(e.g., McLeod \& Rieke 1995; Heraudeau et al. 1996; Mulchaey \& Regan
1997; Ho, Filippenko, \& Sargent 1997) or to have recently experienced
a major interaction or merger (Fuentes-Williams \& Stocke 1988;
Dultzin-Hacyan 1998; De~Robertis, Yee, \& Hayhoe 1998; Virani, de
Robertis, \& Van Dalfsen 2000; although see last paragraph of this
subsection). These results seem to favor minor intrinsic processes
over large-scale external processes for the fueling of low-luminosity
AGNs. Ejecta from a nuclear star cluster (e.g., Norman \& Scoville
1988; Petty 1992; Williams, Baker, \& Perry 1999) may be all that is
needed in the cases of Seyferts and other low-luminosity AGNs to keep
their nuclei active. The ``angular momentum problem'' in feeding
low-luminosity AGNs may therefore reduce to forming the dense stellar
cluster in the first place. Nuclear starbursts are the prime
candidates for the formation of these clusters.

Several studies have shown that the molecular material needed to fuel
nuclear starbursts in Seyfert galaxies is present near the nuclei of
these objects (e.g., Meixner et al. 1990; Tacconi et al. 1997; Kohno
et al. 1998; Baker \& Scoville 1998).  But is this material forming
stars?  Direct evidence for recent {\em nuclear} star formation now
exists in a number of Seyfert 2 galaxies (i.e. Seyferts without broad
recombination lines). Optical and ultraviolet spectroscopy of the
nuclear regions of these galaxies often reveals the signatures of
young and intermediate-age stars.  The stellar Ca II triplet feature
at $\lambda\lambda$ 8498, 8542, 8662 in Seyfert 2s has an equivalent
width similar to that in normal galaxies while the stellar Mg~Ib
$\lambda$5175 is often weaker (Terlevich, Diaz, \& Terlevich
1990). This result is difficult to explain with a combination of an
old stellar population and a featureless power-law continuum from an
AGN. The most natural explanation is that young red supergiants
contribute significantly to the continuum from the central
regions. Evidence for intermediate-age (a few 100 Myrs) stars in these
galaxies is also apparent in the blue part of the spectrum, where the
high-order Balmer series and He I absorption lines appear to be
present in more than half of the brightest Seyfert 2 galaxies (e.g.,
Cid Fernandes \& Terlevich 1995; Gonz\'ales Delgado, Heckman, \&
Leitherer 2001). A few of these objects may even harbor a broad
emission feature near 4680 \AA, possibly the signature of a population
of young (a few Myrs) Wolf-Rayet stars (Gonz\'ales Delgado et
al. 2001).  The ultraviolet continuum from some of the brightest UV
Seyfert 2s also appears to be dominated by young stars based on the
strength of absorption features typically formed in the photospheres
and in the stellar winds of massive stars (e.g., Heckman et al. 1997;
Gonz\'ales Delgado et al. 1998). The extended, soft, thermal X-ray
emission from these objects seems to confirm these results (Levenson,
Weaver, \& Heckman 2001). The bolometric luminosities of these nuclear
starbursts ($\sim$ 10$^{10}$ L$_\odot$) are similar to the estimated
bolometric luminosities of their obscured Seyfert 1 nuclei.

Interestingly, a distinction appears to exist between Seyfert 1s and
Seyfert 2s. Seyfert 2 galaxies have long been known to present a
larger far- and mid-infrared excess than Seyfert 1s (e.g.,
Rodriguez-Espinosa, Rudy, \& Jones 1986; Dultzin-Hacyan et al. 1988;
Pier \& Krolik 1993; Maiolino et al. 1995), but most of this excess
emission may be attributed to star formation in the host galaxy rather
than from a nuclear starburst.  The departure of the galaxy or bulge
blue luminosity of Seyfert 2s from the Tully-Fisher and Faber-Jackson
relationships (e.g., Whittle et al. 1992a, 1992b; Nelson \& Whittle
1995) and the diffuse radio emission around some Seyfert nuclei
(Wilson 1988) may have the same origin. However, recent investigations
have also suggested excess {\em nuclear} starburst activity in Seyfert
2s relative to Seyfert 1s (Gonz\'alez Delgado et al. 2001; Gu et
al. 2001) and possibly a higher frequency of companions near type 2
objects (e.g, De Robertis et al. 1999; Dultzin-Hacyan et
al. 1999; Levenson, Weaver, \& Heckman 2001). These results cannot be explained in the context of the
Seyfert unification theory (which purports that Seyfert 1s and 2s are
basically the same type of objects seen from different perspectives),
but they may reflect an evolutionary connection between starbursts,
Seyfert 2s, and Seyfert 1s.

\subsection{High-Luminosity Regime: QSOs, FR II Radio Galaxies, ULIRGs}

The stringent requirements on the mass accretion rates for luminous
AGNs almost certainly require external processes such as ``major''
galaxy interactions or mergers to be involved in triggering and
sustaining this high level of activity over $\sim$ 10$^8$
years. Substantial evidence exists that the precursors to at least
some powerful AGNs have indeed been gas-rich mergers.  Classical
double (FR II) radio galaxies have long been known to show tidal tails
and other signs of interaction (Smith \& Heckman 1989; Baum, Heckman,
\& van Breugel 1992).  Evidence for recent or on-going galactic
interactions is also seen in several quasars (e.g., Hutchings et
al. 1994; Bahcall et al. 1997; Boyce et al. 1998).  Abundant molecular
gas has been detected in radio galaxies and quasars (e.g., Sanders et
al. 1988b, 1989b; Barvainis et al. 1989, 1995; Mirabel, Sanders, \&
Kaz\'es 1989; Scoville et al. 1993; Ohta et al. 1996; Omont et
al. 1996; Evans et al. 1999a,b), and many of them also show the
spectroscopic signatures of recent star formation (e.g., Tadhunter,
Dickson, \& Shaw 1996; Tran et al. 1999; Brotherton et al. 1999). The
far-infrared excess in some of these objects may also be attributed to
star formation (e.g., Rowan-Robinson 1995; see Sanders et al. 1989a
for another interpretation). In this merger scenario, quasars were
more common in the past because of the enhanced frequency of
collisions [$\propto$ (1+z)$^{4.0 \pm 2.5}$; e.g., Zepf \& Koo 1989]
and the larger proportion of unprocessed gas. Moreover, the observed
redshift cut-off for quasars ($z$ $\approx$ 5) marks the epoch at
which disk systems formed.

ULIRGs may provide the clearest observational link between galaxy
mergers, starbursts and powerful AGNs.  Nearly all ULIGs show strong
signs of advanced tidal interactions (e.g., Sanders et al. 1988a;
Melnick \& Mirabel 1990; Murphy et al. 1996; Clements et al. 1996).
All of them are very rich in molecular gas (e.g., Solomon et al. 1997;
Frayer et al. 1998, 1999), most of which is distributed well within
the inner kpc of the galaxy (e.g., Downes \& Solomon 1998; Bryant \&
Scoville 1999; Sakamoto et al. 1999). They also present a large
concentration of activity in their nuclei, including strong optical
emission lines characteristic of a starbursting stellar population and
in about 30\% of cases, broad or high-ionization emission lines that
suggest the presence of a powerful AGN coexisting with the starburst
(e.g., Kim et al. 1998; Wu et al. 1998a,b; Veilleux et al. 1995,
1999a; Kewley et al. 2001). Similar results are found in the
near-infrared (e.g., Goldader et al. 1995, 1997; Veilleux et al. 1997,
1999b) and in the mid-infrared (e.g., Genzel et al. 1998; Lutz et
al. 1998; Rigopoulou et al. 1999; Lutz, Veilleux, \& Genzel 1999;
Dudley 1999)

The fraction of AGN-dominated ULIRGs is significantly larger among
objects with high infrared luminosities and warm infrared colors (e.g.,
Kim et al. 1998; Veilleux et al. 1995, 1999a,b; Wu et al. 1998b;
Kewley et al. 2001). Current results on a limited set of ULIRGs (e.g.,
Rigopoulou et al. 1999) suggest that the dominance of AGN or starburst
in ULIRGs may depend on local and short-term conditions (e.g.,
compression of the circumnuclear interstellar medium as a function of
gas content and galaxy structure, local accretion rate onto the
central black hole, etc.) in addition to the global state of the
merger.  Still several lines of evidence suggest that {\em warm}
ULIRGs are indeed more advanced, transition objects and that (radio
quiet) QSOs correspond to the final state of the merger-induced
sequence ``starburst $\rightarrow$ ULIRGs $\rightarrow$ QSOs'' (e.g.,
Surace \& Sanders 1999; Scoville et al. 2000; Zheng et al. 1999;
Veilleux et al. 2001).

\section{Unanswered Questions}

The exact nature of this starburst-AGN connection is not at
all clear.  Unanswered questions include:

\begin{itemize}

\item[\bf 1.] Can an AGN be triggered without a burst of star formation?
\item[\bf 2.] If a SMBH is indeed present in every (massive) galaxy, can 
a starburst be taking place without any AGN activity?
\item[\bf 3.] Can starbursts and AGNs simply coexist without interacting with each other?
\item[\bf 4.] If not, in what way are the starbursts and AGNs interacting?
\subitem Is mass loss from the central stellar cluster fueling (low-luminosity) AGNs?
\subitem Is the molecular gas ``unused'' by the starburst feeding the SMBH?
\item[\bf 5.] Is there an evolutionary connection between starbursts and AGNs? 
\subitem Which of the starburst or AGN comes first? 
\subitem Is the merger-induced starburst $\rightarrow$ QSO model correct?
\subitem Is there a similar sequence in low-luminosity objects? 
\subitem Is there a evolutionary connection between narrow and broad-line objects?
\item[\bf 6.] Is the nature of the starburst-AGN connection dependent on look-back time? 

\end{itemize}

Some of these questions should be testable in the near future. For
instance, in the merger-induced scenario starburst ages should
increase along the sequence ``starburst $\rightarrow$ cool ULIRGs
$\rightarrow$ warm ULIRGs $\rightarrow$ quasars''. Detailed
spectroscopic studies should be able to answer this
question. Increasingly sensitive techniques and instruments to detect
obscured AGNs (e.g., infrared and X-ray spectroscopy from the ground
and with satellites) will allow to put better constraints on the
contribution of the AGN to the total energy output of galaxies
(questions \#2). Questions regarding the starburst-AGN connection in
the early universe will obviously be more difficult to answer. For
these we probably have to wait for the next generation of ground-based
telescopes and astronomical satellites to decipher the nature of the
starburst-AGN connection in proto-galaxies.  In the meantime, much
effort should be invested in {\em predicting} what we should expect to
see!


%

\end{document}